\documentclass[aps,prd,reprint,secnumarabic,nonbibnotes,floatfix,amssymb,longbibliography,showpacs,preprintnumbers]{revtex4-1}

\usepackage{graphicx}
\usepackage{subfigure}
\usepackage{mathptmx}
\usepackage{helvet}
\usepackage{lineno}
\usepackage{textcomp}
\usepackage{multirow}
\usepackage{amsmath}

\begin{document}

%\title{ Study of the Lund string with a helix structure}
\title{ Soft QCD path to the mass hierarchy}

\author{\v{S}\'{a}rka Todorova-Nov\'{a}}  
\affiliation{LAPP, Annecy, France}
\email[sarka.todorova@cern.ch]{}

\begin{abstract}   
   
   The study of the quantization of the QCD string with the helix structure ~\cite{hel2}    
   is put into the context of a decades-long discussion opposing the
   probabilistic and the deterministic interpretations of the quantum
   theory. The recent evolution of the string fragmentation model (to
   a large extent driven and confirmed by the empirical evidence),
   towards the deterministic point of view is recounted: the notion of causality not only paves
   a way to the study of mass spectrum, it also resolves  
   a long-standing ambiguity about the nature of the so-called Bose-Einstein correlations
    ~\cite{chains}.  Two directions for the further development of the model are
    outlined: 1/ the elimination of soft collinear divergencies,  2/ the relation between
   the topological properties of the QCD string and the emergence of new
   particle types (quantum numbers). 
  
\end{abstract}

\pacs{13.66.Bc,13.85.Hd,13.87.Fh}

\maketitle

%\input{introduction.tex}
%\input{pt_modelling.tex}
%\input{observables.tex}
%\clearpage
%\input{charged_assymetry.tex}
%\input{overview.tex}
%\input{conclusions.tex}
%\input{appendixA.tex}

\section{Introduction}

     The absence of predictions concerning the particle mass is
    a commonly accepted feature of the probabilistic quantum
    theory. The fact that there is a possibility to develop a quantum model which does
    provide a key to the mass spectrum therefore may come as a
    surprise, in particular when the novel feature stems directly from the
    integration of the causality constraint in the model.  It is also
    unusual to start a conventions-breaking model development in the domain of the
    non-perturbative QCD, considered too complex and out of bounds for
    reliable calculations. Yet this is what has happened in the past
    decade,  practically unnoticed in the shadow of the highly publicized
    hunt for Higgs boson(s).  

\section{The origins}
     Experimental high energy physics relies heavily on the
    semi-classical model of QCD string fragmentation to describe the formation
    of hadrons and of the hadronic jets observed in the data. The
    Lund string fragmentation model ~\cite{lund}, at the core of a widely used
    Monte-Carlo event generator Pythia ~\cite{pythia}, is so
    successful in this exercise that the modelling is often taken for
    granted; model development is replaced by tuning of
    parameters associated with various subcomponents of the
    modelling.  In case the underlying physics picture for some of the
    subcomponents is inadequate, a systematic untunable discrepancy
    is observed in the comparison of the model with the data;
    due to limited number of observables, there is however
    competition between contributions from different subcomponents
    and it becomes difficult to identify the main source of the discrepancy.  
    Progress in the modelling is therefore relatively slow under
    the best of circumstances; once the precision of the modelling of the hadronization
    becomes acceptable for the subsequent analysis, and in the
    absence of new ideas, the model becomes practically frozen. This
    is precisely what happened to the Lund string fragmentation
    during the LEP era: the model has been put through extensive
    tuning on the data samples from the hadronic Z0 decay; minor
    changes have been introduced but the original picture of a
    1-dimensional string fragmentation completed with a tunnelling
    process to generate intrinsic transverse momentum of hadrons has
    been preserved. It has been duly noted that the performance of the 
    model in the transverse region is poor ~\cite{z0_DELPHI} but
    the tunnelling process has not been particularly blamed for it 
    (the largest discrepancies appear in the tail of $p_T$
    distributions ).  It is perhaps time to reconsider this attitude : 
    at LEP energies, for $p_T<$ 1 GeV,  the precision of the modelling
    is well below 10\% for LEP data ;  at LHC-Run1 energies, the discrepancies
    reach 20-30\% in the same region. As far as the intrinsic hadron
    transverse momentum is concerned, are there alternatives to 
    tunnelling ?   

\section{Helix-shaped QCD string and LEP data}   
   
       The first and most important step in the replacement of the
     tunnelling has been suggested by Andersson et al
     ~\cite{lund_helixm}.  The study has been primarily motivated by
     the question of regularization of the end of the parton shower:
     on the basis of the requirement of  helicity conservation
     in the gluon emission, and as a result of the study of optimal
     packing of soft gluons in the phase space, the authors concluded that
     the QCD field created by a pair of colour connected partons should acquire a helical shape and, as a
     consequence, a non-zero transverse extension.  The observation
     that the gluon emission cannot proceed via a collinear gluon
     emission and that the consecutive emissions are probably ordered
     in azimuthal angle has several  highly non-trivial
     implications:
     \begin{itemize}
     \item{ the emitting/absorbing parton} (quark) 
     acquires an effective mass and angular momentum ( the cyclical interaction
     with the field becomes a mass generating mechanism )
     \item{ in the fragmentation process,}  the string
     tension operates in 3 dimensions and generates the intrinsic
     transverse momentum of hadrons (replacing tunnelling)
     \item{the helix shape of the string} is translated into correlations
     between transverse and longitudinal components of the hadron momenta.
     \end{itemize}
      
     The azimuthal ordering of hadrons ( following the shape of the
     helical string ) and the correlations between longitudinal and
     transverse components of hadron momenta should be experimentally
     observable, as well as modifications of the inclusive $p_T$
     spectra. There is however a certain ambiguity  concerning the shape of
     the helix : the phase-space distance between soft gluons forming
     the colour chain has both longitudinal and transverse  components
      - in $\Delta y$ (rapidity)  and $\Delta
     \phi$ (azimuthal angle) . The initial search for evidence of
     azimuthal ordering of hadrons has been done for a helix
     winding proportional to the rapidity difference between hadrons
     and no clear signal has been observed ~\cite{delphi_scr}. The
     comparison of an alternative helix parametrization ( with winding
     density proportional to the energy density of the string ) has
     been more successful - it has been shown that the helix fragmentation significantly
     improves the agreement with the data in observables based on the transverse momentum
     ~\cite{tuning}. The most significant improvement occurs in $<p_T>
     vs. x_P$ ( average $p_T$ as function of scaled momentum ) which
     measures precisely the sort of correlation which distinguish the
     helical string based fragmentation from uncorrelated transverse
     momentum generation in the standard Lund string fragmentation. 
     The study of inclusive spectra in hadronic Z$^0$ decays at LEP is a first, admittedly indirect, evidence in favour of the
     model of helical QCD string.
         
\section{Helix-shaped QCD string and LHC data}   
   
     A further piece of evidence has been provided by the ATLAS
     Collaboration  which found evidence of azimuthal  ordering of
     hadrons in minimum bias events with minimal acollinear jet
     activity ~\cite{atlas_ao}. The effect can be partially reproduced
     by a modified Pythia fragmentation routine incorporating the
     helix string shape ~\cite{pystrf}.  The size of the measured
     effect is larger than the one expected for direct hadrons - the
     data suggest the decays of resonances carry, to some extent, the memory of the
     generating string field.  It seems straightforward to extend
     the fragmentation model to include the decays of short-lived resonances
     which can be viewed as a continuation of the string fragmentation
     process, but such an extension destroys the good agreement
     between the model and the LEP data in the low $p_T$ region: the
     model produces too many soft hadrons.  There seems to be some kind
     of a natural cut-off in the production of soft hadrons which can
     be related to quantum effects which are not explicitly included
     in the model.  Somewhat unexpectedly, the solution to the
     problem comes from the analysis of the space-time properties of
     the helical string.

\section{Helix-shaped QCD string and causality}  
  
      The ``folding'' of the 1-dimensional QCD string into a
      3-dimensional object  brings a new feature into the modelling :
      the possibility to create a causal connection between adjacent string breakups ( which are strictly
      causally disconnected in the standard Lund string fragmentation ). 
      It seems to be an excellent idea on its own to integrate the cross-talk
      between breakup vertices into the model but the result by far exceeds the expectations : it
      literally propulses the model on a qualitatively new level. It is somewhat difficult to
      understand how a narrow resonance can emerge from random uncorrelated
      string breakups in the standard Lund string fragmentation; with
      the causality restored, the quantization pattern describing the
      light mesons appears rather naturally \cite{hel2}.  Under the
      causality constraint, the longitudinal momentum decouples from
      the transverse mass of hadrons which becomes the quantized
      quantity ( the mass of hadron depends on the transverse
      properties of the string only ).  The parameters describing the
      shape of the QCD field can be extracted from the mass spectrum
      of light pseudoscalar mesons with a precision of $ \sim 3 \% $. Furthermore, the
      transverse size of the string obtained through this fit agrees
      with estimates obtained from the fit of the glueball spectrum
      in topological QCD ~\cite{buniy}. However, the most stringent
      evidence for the helical winding of the string comes from the
      analysis of 2-particle correlations.

 \section{Helix-shaped QCD string and 2-particle correlations}

      The quantization rule derived from the spectrum of light mesons
      defines the ground state hadron (pion) as a piece of helicoidal
      string with phase difference of $\Delta \Phi \sim 2.8$ rad. If a
      helix string with regular, or slowly varying winding fragments
      into a chain of ground state charged hadrons, the model predicts
      emergence of a correlation pattern characterized by the excess of
      close like-sign pairs: the quantum threshold for the production 
      of adjacent (unlike-sign) hadrons depletes the low Q region
      while a sort of Bose-Einstein condensate emerges from
      the association of closest like-sign particles ( with phase separation of 
      2 times 2.8 rad $\sim$ 0.7 rad $\ll$ 2.8 rad).  The chain of n ground state
      hadrons is characterized by mass 
 \begin{equation}
    m({\rm n } ) \leq {\rm n }  \  0.19  \ {\rm GeV}.
\end{equation}

      The analysis of ATLAS minimum bias events ~\cite{chains} finds
      a very good agreement between the data and the model
      predictions; it seems the so-called Bose-Einstein correlations
      stem entirely from the coherent production of ground state
      hadrons. Thus the deterministic variant of the helix string
      fragmentation accomplishes what several decades of intensives
      studies of the effect did not achieve - it discards the hypothesis of
       incoherent origin of correlations, and provides a recipe to
      reproduce the effect in MC models, with parameters fully constrained
      by the fit of the mass spectrum of  light mesons.

\section{Helix-shaped QCD string and the quantum theory of motion}

       Profound similarities can be found between the quantized
     helix string model and the quantum theory of motion
     \cite{holland}.  The quantum potential of the later
     corresponds to the effective transverse string shape; the
     helix string model takes a step further and associates the
     quantum potential with the interaction of the charged particle with
     the field quanta . Most of the argumentation developed in
     \cite{holland} applies directly to the helix string model. In
     particular, it has been argued that the quantum potential
     defines the non-classical but unique particle trajectory satisfying
     the principle of uncertainty.
\begin{figure}[hbt!]
\begin{center}
\includegraphics[width=0.3\textwidth, angle = 90]{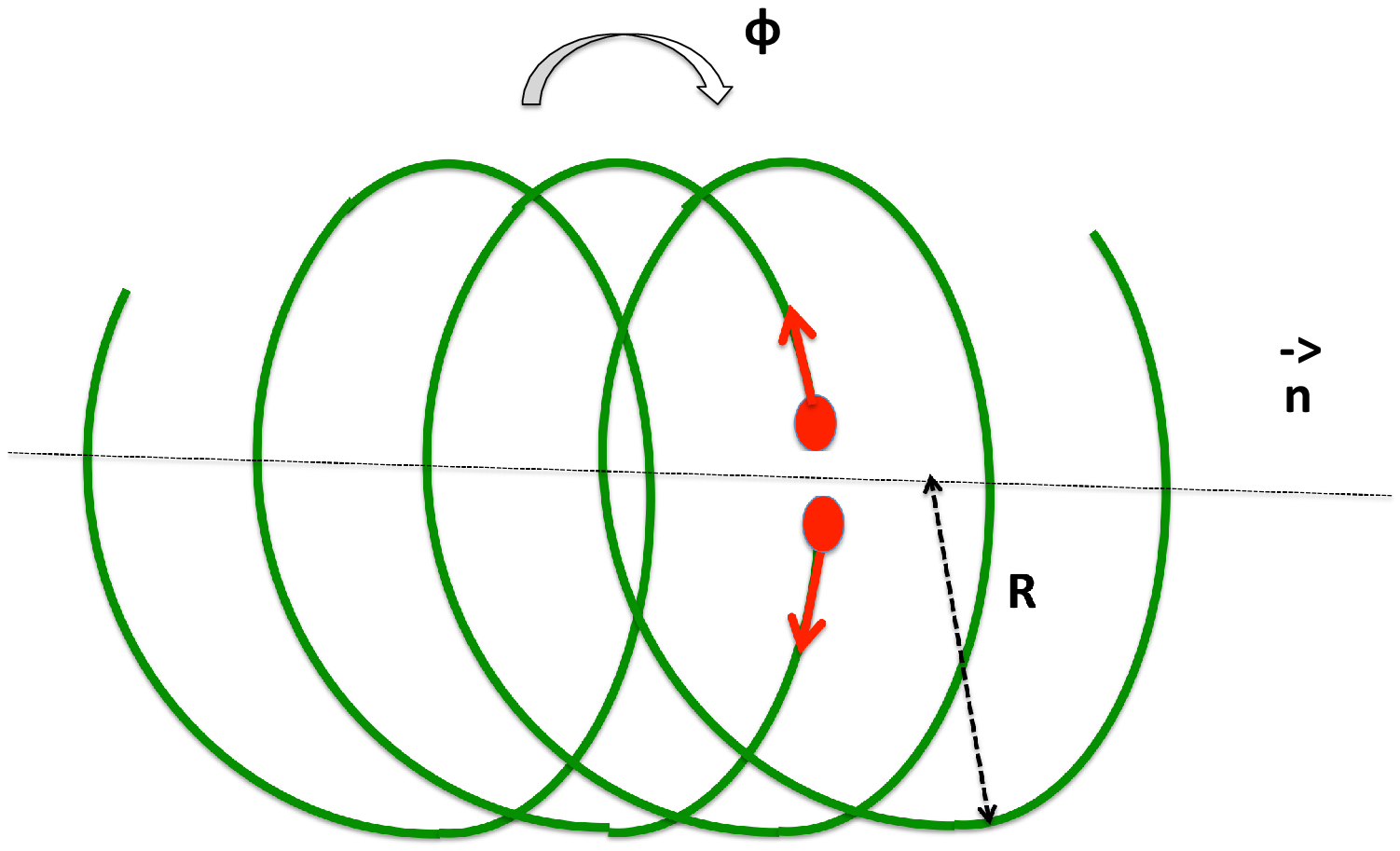}
\caption{ 
 The breakup of the helix-shaped string.
\label{fig:helix}}
\end{center}
\end{figure}

      Let's look in more detail on the parton trajectory in the
    quantized helix string model. Imagine the helical string breaking
    by the $g \rightarrow q\bar{q}$ process (Fig.~\ref{fig:helix}); the quarks
    are treated as massless and their initial momentum is considered negligible.
    Under the action of the string tension ( surrounding field )
    the quarks follow the curved trajectory which is non-classical : they are spinning around
    the direction of string axis acquiring ``macroscopic''  longitudinal momentum, angular momentum and effective
    mass; the average transverse momentum is vanishing.
    Let's set the starting parton configuration at t=0, with string
    axis passing through the origin and pointing along $\vec{n}$.
    After a time interval $\Delta t$, the position of the parton becomes
    \begin{equation}
      x(\Delta t) = [ \Delta t,\Delta t \beta c \vec{n} , \vec{R} \exp^{\i \omega \Delta t}] 
    \end{equation}

     where vector $\vec{R}$ points in the direction of the helix trajectory at the origin ( breakup point ), and
    $\vec{n} \cdot \vec{R}$=0; the momentum acquired by the parton via the interaction with the string (field) is
        
    \begin{equation}
      p(\Delta t) = \kappa c [ \Delta t,\Delta t \beta c \vec{n} , \vec{R} (\exp^{\i \omega \Delta t} - 1) ] 
    \end{equation}
     
     where $\kappa$ stands for the string tension, and $\beta = \sqrt{
       1 - (R\omega/c)^2 }$. 
%The helicity factors in the exponent are neglected for the moment.

    The particle appears on the classical trajectory (= measured position + $\beta c \Delta t \vec{n}$) 
    just once per the period, for $\omega \Delta t$ = k 2$\pi$, k=1,2,.... The corresponding string action can be calculated
    (k=1):

   \begin{equation}
      x(\Delta t) \cdot p(\Delta t)  - x(0) \cdot p(0)  = \kappa c \frac{(\Delta t)^2}{\gamma^2}
      = \kappa c \tau^2
   \end{equation} 

   where $\tau = \frac{2\pi}{\omega\gamma}$ stands for the invariant period of parton rotation (spin).

   It is possible to calculate $\tau$ from the radius of the helix string ( obtained independently from
  the fit of light pseudoscalar mesons \cite{hel2}, and from the shape of 2-particle correlations \cite{chains}, for $\kappa$ = 1 GeV/fm ):
 
  \begin{equation}
     \tau = \frac{ 2 \pi R }{c} = (0.427 \pm 0.013) \  {\rm fm}/c
  \end{equation}

   which leads us to the estimate 
  \begin{equation}
    \kappa c \tau^2  = \kappa (0.183 \pm 0.01)  \ {\rm fm}^2/c \simeq \hbar.  
 \end{equation}
    
   ( Equivalent result is obtained from the integral of the string action using the Lagrangian constructed
  from the field potential and the kinetic energy of the parton ). 

\section{Helix-shaped QCD string and the spin of partons}
   
    We arrive at the point where the quantized helix string model (more specifically, the accumulated 
 empirical evidence accompanying its development) does not leave other choice but to associate the angular
 momentum stemming from the interaction with the field quanta with the spin of the emitting (absorbing) parton.
% ( If we want to profit from the regularization of the soft hadron production via the quantum threshold, and
 %from the predictions concerning the 2-particle correlations, we need to accept the existence of observable
 %angular momentum associated with the curved trajectory).
  
   As stated in the introduction, it is highly unusual to attempt to develop the spin model in the domain of non-perturbative
  QCD. It is nevertheless possible that the study of hadronization is where the problematics becomes most accessible from
  the experimental point of view, due to the trace of the ordered gluon emission preserved by the confinement.  
   
    The interpretation of the spin in terms of gluon and photon
    emission/absorption can actually help to understand some
    non-intuitive properties of the spin.

    As explained in Section III, the helicity conservation requires
    that the emitting parton and the emitted
    gluon have to go apart. In the transverse plane, the emission
    is equivalent to a phase change along the effective helix-like
    trajectory of the emitting quark $\phi \rightarrow \phi + d\phi $.
    The emitted gluon acquires transverse momentum and phase
    \begin{eqnarray}
      p_{T}^g &= & 2 R |sin(d\phi/2)| \\
     \phi^g &= & \phi + d\phi/2  
   \end{eqnarray}          

     and the quark, in virtue of momentum conservation, carries the
     recoil transverse momentum $p_{T}^g exp^{i \phi^g+\pi}$. 
  
     In a purely mathematical approach, fancy effects can be obtained
     with spinors using $d\phi > \pi $, in particular $d\phi = 2\pi $
     which seemingly reverts the parity of the particle state.  The physics
     case of the gluon or photon emission is different : the range
     of the possible phase change for a single emission is limited by  $|d\phi|\leq \pi$,
     with positive/negative values distinguishing the
     positive/negative helicity (right-handed and left-handed
     emission). It takes at least two subsequent emissions (or an
     emission and an absorption)  of field quanta to accomplish
     a phase change of 2$\pi$ for the emitting particle, for which
     the recoil vanishes; the result is nevertheless not an identity
     since the particle moves along the longitudinal direction of the
     string, too.

\begin{figure}[ht!]
\begin{center}
\includegraphics[width=0.5\textwidth]{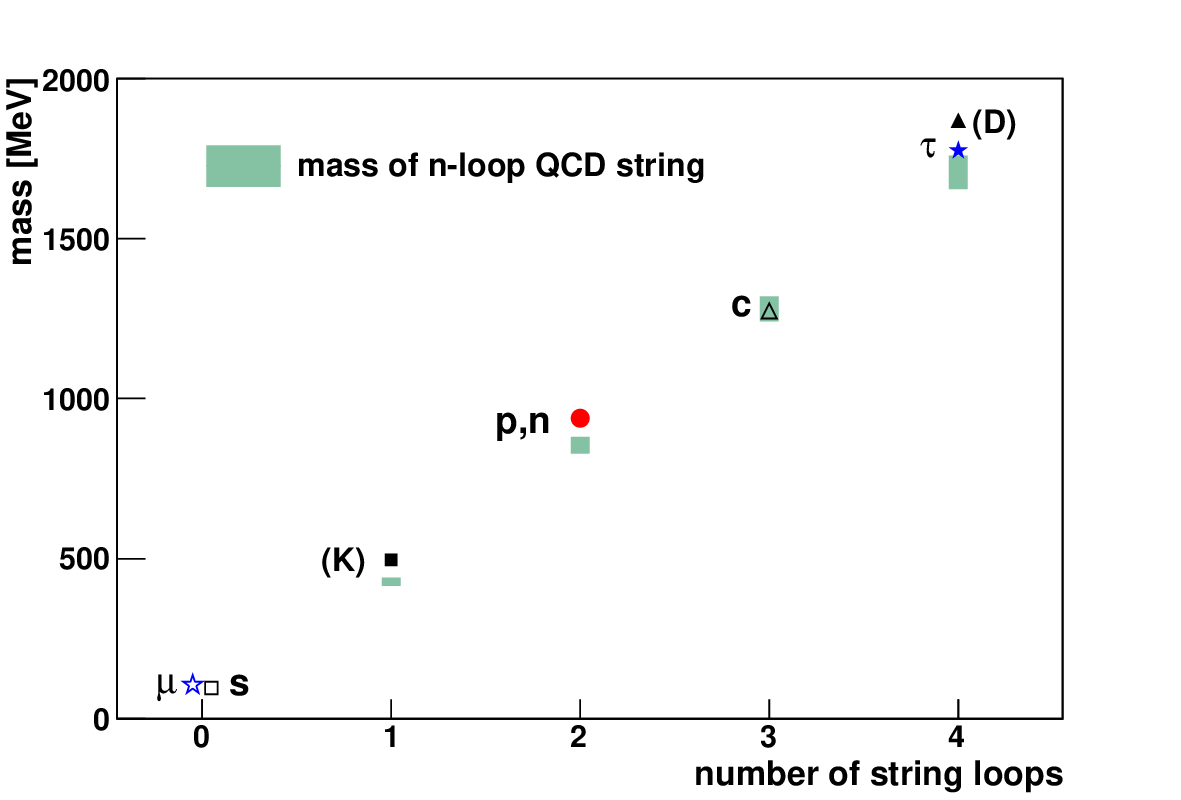}
\caption{ 
 The mass spectrum of particles \cite{pdg} carrying a new quantum number
 compared with the mass of a helix-shaped QCD string with parameters
 derived from experimental data. The open and bound quark states are
 included  (s-K, c-D). 
\label{fig:knots}}
\end{center}
\end{figure}

   \section{ Helix-shaped QCD string and the emergence of new quantum
     numbers } 
  
      In \cite{hel2}, the mass spectrum of light mesons emerging from
      the quantization pattern of the helix-shaped QCD string has been
      discussed.  But there seems to be even deeper connection between
      the topology of the helix-shaped QCD string and the mass
      spectrum : the new particle types (quantum numbers) seem to
      emerge in mass interval which corresponds to the mass of a
      single loop of the  helix-shaped string.
      The mass stored in 1 loop of the helical QCD string is, according to the
      current best estimates,
          
      \begin{equation}
        m ( {\rm 1 \ loop} ) = 2 \pi \kappa R = (0.427 \pm 0.013) {\rm
          GeV }.
      \end{equation}

       Fig.~\ref{fig:knots}  shows the comparison of the mass of
       N-loops with the mass of (lightest) particles carrying a new quantum number.
  
       The picture suggests these quantum numbers are related to the 
       topological knotting of the string across the loops with a
       binding energy of  $\sim$ 70-100 MeV.  
 
 \section{ Helix model and the elimination of soft collinear
   divergencies }

     The phenomenological success of the effective quantization as
     described in previous sections is intriguing. It seems the ideas
     advanced by the authors of the helix string model, namely
     the existence of a physical mechanism in the emission of field
     quanta which forbids collinear emission, are correct. The precise
     knowledge of this mechanism is a key ingredient for the
     elimination of soft collinear divergencies from QCD calculations.
     It is not yet clear at this stage if better understanding of this
     problem requires further experimental evidence.  

\begin{figure}[ht!]
\begin{center}
\includegraphics[width=0.25\textwidth,angle=90]{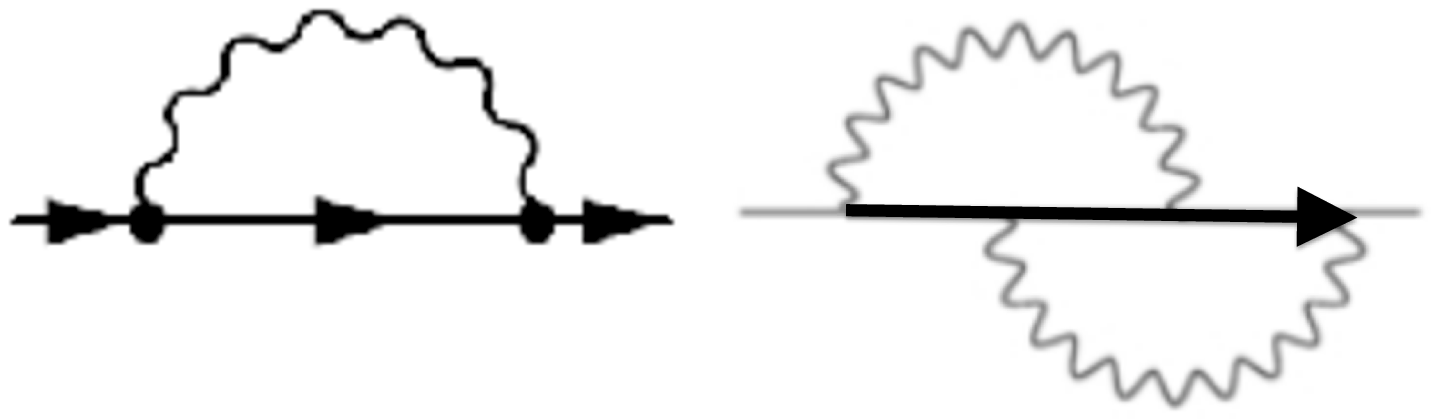}
\caption{ 
 The interaction of particle with the field obeys quantization rules
which are not yet fully understood. The concept of particle
trajectory (discarded by quantum mechanics) should be reexamined.  
\label{fig:self-energy}}
\end{center}
\end{figure}

      It s worth noticing that the arguments concerning the interaction of charged
      particles with the field are valid for the electroweak
      processes, too. It is legitimate to ask which effective mass
     a free electron acquires by the interaction with its own photon
     cloud. The study of properties of the helical QCD string suggest
     the photon emission may be strictly ordered and the momentum
    circulating within loops in Fig.~\ref{fig:self-energy} quantized.
     The question seems to be closely related to the problem of vacuum
     energy, too : the quantization  of loop
     calculations in the frame of a deterministic quantum theory
     does not suffer from soft divergencies inherent to the
     quantum mechanical approach. 
   \section{Final remarques}
         The model of helix-shaped QCD string  provides a bridge
         between several seemingly disconnected areas of research in
         the particle physics ( mass spectrum of hadrons, spin
         physics, particle correlations ).
It is doing remarkably well
      is some of notoriously problematic domains of the particle
      production (description of intrinsic hadron transverse
      momentum, Bose-Einstein correlations).  In the frame of the
      model, a close relation between causality and quantum properties
      can be established. The resulting phenomenological construction
      is in very good agreement  with experimental measurements. 
     There is a large variety of complementary measurements which
     can be performed in order to test the model: verification of the
     prediction of a $p_T$  threshold in the production of direct
     hadrons, study of the asymmetry of particle correlations with
     respect to the orientation of string, the link between azimuthal
     ordering and the study of ordered hadron chains, polarization
     studies etc.  In the author's opinion, there is already enough
    experimental evidence to demonstrate the power of the
    deterministic approach to the quantum effects and its superiority
    with respect to the purely probabilistic interpretation of the
    quantum physics, but it is also clear there is still much work to
    do to reconciliate the physics community on this subject.
      Two areas of possible further development of the model from the
      phenomenological point of view are outlined (the regularization
      of soft emissions and the
      topological interpetation of mass hierarchy).

\bibliography{draft}

\end{document}